\documentclass[letterpaper]{article} 
\usepackage{aaai2026}  
\usepackage{times}  
\usepackage{helvet}  
\usepackage{courier}  
\usepackage[hyphens]{url}  
\usepackage{graphicx} 
\usepackage{natbib}  
\usepackage{caption} 
\usepackage{algorithm}
\usepackage{algorithmic}

\usepackage{url}
\usepackage{pifont}
\usepackage{xcolor}
\usepackage{epstopdf}
\usepackage{subfigure}
\usepackage{subcaption}
\usepackage{multirow}
\usepackage{booktabs}      
\usepackage{amsmath}
\usepackage{amssymb}

\usepackage{newfloat}
\usepackage{listings}
\DeclareCaptionStyle{ruled}{labelfont=normalfont,labelsep=colon,strut=off} 
\floatstyle{ruled}
\newfloat{listing}{tb}{lst}{}
\floatname{listing}{Listing}
\title{AutoGameUI: Constructing High-Fidelity GameUI via Multimodal Correspondence Matching}
\author{
    Zhongliang Tang\equalcontrib,
    Qingrong Cheng\equalcontrib,
    Mengchen Tan\equalcontrib,
    Yongxiang Zhang,
    Fei Xia\thanks{Corresponding author.}
}
\affiliations{
    TiMi L1 Studio, Tencent, China\\
    \{clarezltang, kirolcheng, mengchentan, freizhang, wallacexia\}@tencent.com
}

\usepackage{bibentry}

\begin{document}

\maketitle

\begin{abstract}
Game UI development is essential to the game industry. However, the traditional workflow requires substantial manual effort to integrate pairwise UI and UX designs into a cohesive game user interface (GameUI). 
The inconsistency between the aesthetic UI design and the functional UX design typically results in mismatches and inefficiencies. To address the issue, we present an automatic system, $\textbf{AutoGameUI}$, for efficiently and accurately constructing GameUI. The system centers on a two-stage multimodal learning pipeline to obtain the optimal correspondences between UI and UX designs. The first stage learns the comprehensive representations of UI and UX designs from multimodal perspectives. The second stage incorporates grouped cross-attention modules with constrained integer programming to estimate the optimal correspondences through top-down hierarchical matching. The optimal correspondences enable the automatic GameUI construction. We create the GAMEUI dataset, comprising pairwise UI and UX designs from real-world games, to train and validate the proposed method. Besides, an interactive web tool is implemented to ensure high-fidelity effects and facilitate human-in-the-loop construction. Extensive experiments on the GAMEUI and RICO datasets demonstrate the effectiveness of our system in maintaining consistency between the constructed GameUI and the original designs. When deployed in the workflow of several mobile games, AutoGameUI achieves a 3$\times$ improvement in time efficiency, conveying significant practical value for game UI development.
\end{abstract}

\section{Introduction}
With the rise of personal computers and smart mobile devices, the game industry has generated considerable economic and cultural impacts. However, as a critical step in game development, game UI development still faces many challenges. Although many studies \cite{li2022learning,warner2023interactive,jiang2024graph4gui} explore UI automation for web pages or mobile apps, few discover and address the issues in game UI development.

\begin{figure}[t]
  \centering
  \includegraphics[width=1.0\linewidth]{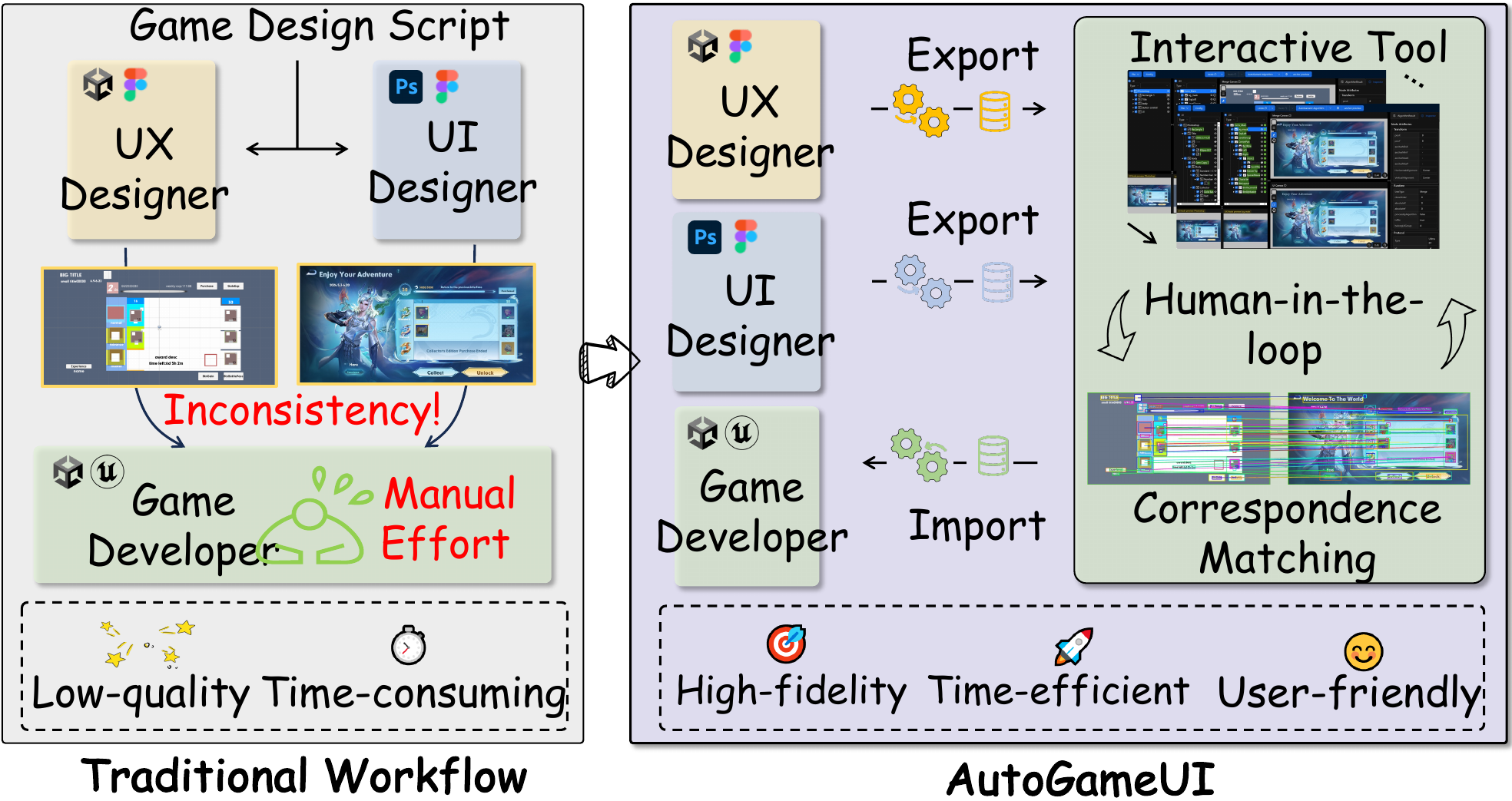}
  \caption{Comparison between the traditional manual workflow and AutoGameUI in game UI development.}
  \label{fig:teaser}
\end{figure}

In the traditional workflow of game UI development, two specialized teams work concurrently to construct a cohesive game user interface (GameUI). The first team, made up of UI designers, focuses on crafting visually appealing elements to achieve diverse game styles. The second team, composed of UX designers, emphasizes usability and functionality by prototyping effective logical structures. As illustrated in Fig~.\ref{fig:teaser}, the distinct focus areas often lead to conflicting design elements and inconsistent layouts, making it difficult to integrate the aesthetic UI design with the functional UX design. Game developers must employ substantial effort to align these differences, such as manually finding correspondence between the UI and UX designs, painstakingly adjusting the position, size, anchor, and asset assignment for the UX elements one by one. This workflow is time-consuming and low-quality, easily causes delays.

To alleviate the bottleneck posed above, this paper introduces AutoGameUI, an automatic system for GameUI construction. As depicted in Fig.~\ref{fig:teaser} and Fig~.\ref{fig:multimodal_learning}, our system comprises two main components: multimodal correspondence matching and an interactive tool. The multimodal correspondence matching is a two-stage learning pipeline, with the first stage focusing on representing UI and UX designs in a latent space, and the second stage aiming to estimate the correspondences between them. Unlike previous studies \cite{yamaguchi2021canvasvae,bai2023layout,jiang2024graph4gui} that examined the representation of graphical UIs from a limited number of perspectives, our method considers various aspects, such as spatial layout, functional semantics, textual content, hierarchy, and even rendering order. Moreover, some works \cite{wu2023screen,otani2024ltsim} attempted to directly estimate correspondences between pairs of graphical UIs based on one-to-one element similarity. In contrast, we adopt the divide-and-conquer strategy to formulate a hierarchical correspondence matching problem. We utilize grouped cross-attention (GCA) modules to compute one-to-group similarity, which, further combined with a constrained integer programming algorithm, enables us to estimate optimal correspondences efficiently. Following prior works \cite{bacikova2013defining,panayiotou2024smauto}, we design a universal protocol to describe the UI and UX designs, enabling the data stream across different platforms. Using the universal data protocol and the optimal correspondences, we can achieve high-fidelity effects in the GameUI construction. To enhance user experience, we develop an interactive tool that provides intuitive feedback and guidance during the construction process. This tool also supports data annotation, allowing us to build the GAMEUI dataset from real-world games for training and evaluation. In summary, this paper makes the following contributions:
\begin{itemize}
  \item We introduce an automatic system for high-fidelity GameUI construction. This system overcomes the bottleneck of traditional manual workflow and significantly accelerates the game UI development.
  \item We propose a two-stage multimodal learning pipeline that estimates optimal UI/UX correspondences from comprehensive representations, incorporating grouped cross-attention modules and constrained integer programming to enhance efficiency and accuracy.
  \item We implement an interactive tool with a universal data protocol that supports cross-platform applications and facilitates human-in-the-loop construction.
  \item We present the GAMEUI dataset, the first-of-its-kind dataset that contains paired UI and UX designs with rich-structured data and high-quality annotations.
\end{itemize}

\section{Related Work}
\subsection{Representation of UIs}
Representing graphical UIs is crucial for tasks such as UI recognition \cite{li2022learning}, UI completion \cite{jiang2024graph4gui}, UI generation \cite{inoue2023layoutdm}, and UI structuring \cite{wu2021screen}. The essential attributes of graphical UIs include spatial layout, semantics, image texture, textual content, view hierarchy, and rendering order. Most methods primarily focus on spatial layout and semantics but overlook others. LayoutTrans \cite{gupta2021layouttransformer}, LayoutDM \cite{inoue2023layoutdm} and LayoutFlow \cite{guerreiro2024layoutflow} generated UIs based on layout geometry and semantics. Other efforts \cite{he2021actionbert, yamaguchi2021canvasvae,jiang2024graph4gui} incorporated image texture and textual content but fell short in understanding hierarchy and rendering order. Some studies \cite{li2021screen2vec,bai2023layout} modeled hierarchy while rendering order remains unexplored. 

\subsection{Correspondence between UIs}
Identifying corresponding elements between pairs of UIs has been investigated for some time now. Earlier studies \cite{kumar2011flexible,kumar2011bricolage} used probabilistic optimization to establish correspondences for automatic design transfer. \citet{dayama2021interactive}
presented an integer programming algorithm to optimize correspondences and developed an interactive tool for layout transfer. LayoutBlend \cite{xu2022hierarchical} found the optimal correspondences between UIs with maintained hierarchical consistencies. \citet{otani2024ltsim} introduced a criterion to measure the similarity between paired layouts. These works are based on heuristic rules, and have poor generalization in practical use.

Some works further applied deep learning methods for correspondence matching. LayoutGMN \cite{patil2021layoutgmn} introduced a graph neural network to determine the similarity between arbitrary layouts. \citet{wu2023screen} used a transformer model to capture latent representations of layout elements and then infer correspondences between UIs. Most of these methods compute a full similarity matrix between pairwise elements. When the number of elements is large, solving the correspondence problem under additional constraints can become extremely time-consuming.

\subsection{Artificial Intelligence in Game Industry}
AI technology has been widely used in the game industry, involving game-playing bots and procedural content generation. Prior works \cite{vinyals2019grandmaster,ye2020mastering,fan2022minedojo} utilized reinforced AI to develop agents, enhancing collaborative play and NPC control. Generative AI has also been employed to create game assets, such as 2D graphics \cite{batifol2025flux}, 3D models \cite{zhao2025hunyuan3d}, music and voices \cite{takada2023genelive,anastassiou2024seed}, and animations \cite{liu2023emage,chen2024taming}.
\begin{figure*}[t]
  \centering
  \includegraphics[width=1.0\linewidth]{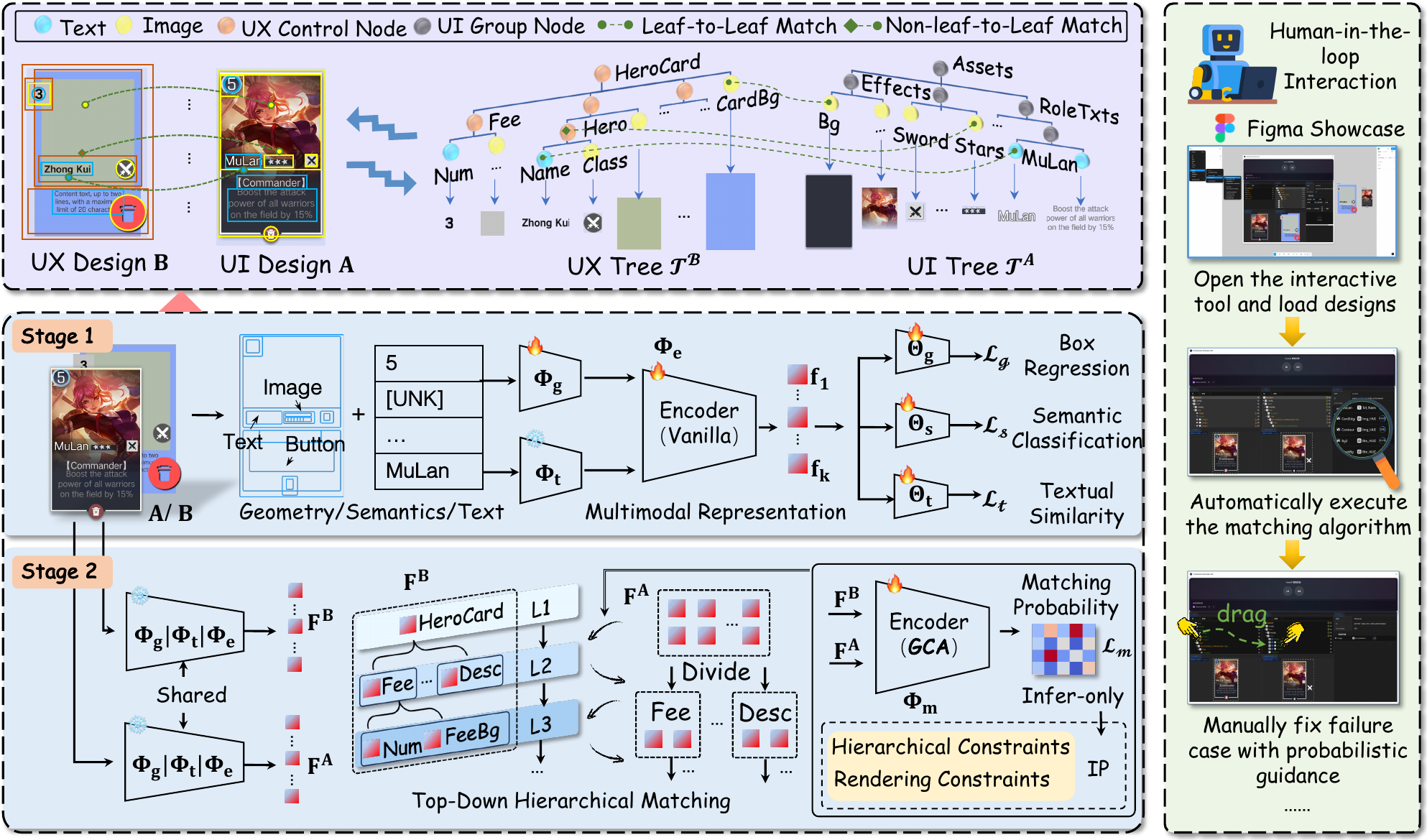}
  \caption{Overview of the AutoGameUI system. Left: multimodal correspondence matching between pairwise UI and UX designs. The first stage focuses on acquiring the multimodal representation of matchable nodes in UI and UX designs. The second stage aims to estimate the optimal correspondences between matchable nodes. Right: interactive showcase in Figma. The user launches the web tool, initiates automatic matching, and manually refines results with matching probabilities.}
  \label{fig:multimodal_learning}
\end{figure*}

Some efforts \cite{li2021understanding,gonccalves2023my} appeared to intersect with human-computer interaction and game development, but their focus has largely been on gameplay and player experience, with limited insights into game UI development. Moreover, intelligent UI development remains in its early stages. Existing tools such as \citet{uizard} and \citet{framer} are primarily designed for web or mobile apps, and perform poorly in real-world game UIs, which often feature aesthetics and complex layouts.

\section{GameUI Construction}
This section outlines our system illustrated in Fig.~\ref{fig:teaser}. We begin by factorizing UI and UX designs into multiple modalities with discrete parameters. Then, we use self-supervised learning to generate continuous multimodal representations and formulate a top-down hierarchical correspondence matching problem. To address it, we introduce the grouped cross-attention module to learn matching probability and apply integer programming for optimal correspondences. With the correspondences, we design a protocol to integrate UI and UX designs. Lastly, we develop a web-based tool to help users interactively build GameUI. 

\subsection{Learning Multimodal Representation}
\subsubsection{Data Attributes} UI and UX designs are typically organized as trees with attributes such as spatial layout, image texture, textual content, hierarchy, rendering order, and semantics. Spatial layout describes the 2D geometry of both leaf and non-leaf nodes, while image texture and textual content are renderable attributes found only in leaf nodes. Unlike previous works \cite{li2022learning,wu2023screen,jiang2024graph4gui}, we exclude image texture, as color differences between UI and UX designs may introduce noise. Hierarchy shows the organization of nodes, with higher-level nodes covering broader scopes. Rendering order determines the display priority of nodes, and semantics indicate node types such as TEXT, IMAGE, or BUTTON.

In this paper, we represent the tree as $\mathcal{T}=(\mathcal{V},\mathcal{E})$, where $\mathcal{V}$ is the set of nodes and $\mathcal{E}$ is the set of edges. Each node $v_k$ is defined as $(g_k, s_k, t_k)$, denoting its 2D geometry, semantic label, and textual content. Each directed edge $e_{ij}$ from $v_i$ to $v_j$ is defined as $(\Delta g_{ij}, \Delta h_{ij}, \Delta r_{ij})$, capturing the relative geometry, hierarchy, and rendering order between nodes. We adopt the identical formula in LayoutGMN \cite{patil2021layoutgmn} for $g_k$, $s_k$, and $\Delta g_{ij}$. Specifically, $\Delta h_{ij}\in\{0,1,2\}$ indicates whether $v_i$ is an ancestor (0), descendant (1), or unrelated (2) to $v_j$ in the hierarchy. $\Delta r_{ij}\in\{0,1,2\}$ indicates if $v_i$ overlaps and is rendered after $v_j$ (0), before $v_j$ (1), or does not overlap (2). 

With the definition, UI and UX designs can be denoted as $\mathcal{T}^{A}$ and $\mathcal{T}^{B}$. The key difference between $\mathcal{T}^{A}$ and $\mathcal{T}^{B}$ is that only leaf nodes are matchable in $\mathcal{T}^{A}$, while both leaf and non-leaf nodes are matchable in $\mathcal{T}^{B}$ (see Fig.~\ref{fig:multimodal_learning}, top). This is because non-leaf nodes in UI are just groups of renderable resources without visual attributes, whereas in UX, non-leaf nodes are controllable widgets or components that can be integrated into GameUI.

\subsubsection{Multimodal Representation}
We utilize a graph neural network $\mathbf{\Phi_{g}}$ \cite{patil2021layoutgmn} to encode the parameters $s_{k}$, $g_{k}$, and $\Delta g_{ij}$. In addition, we use a pre-trained language model $\mathbf{\Phi_{t}}$ \cite{reimers-2019-sentence-bert} to encode the parameter $t_{k}$ and adopt the special token [UNK] to pad nodes without text. These heterogeneous features are projected into a unified feature vector, which is further processed by a transformer encoder $\mathbf{\Phi_{e}}$. With rotary positional encoding \cite{su2024roformer}, $\mathbf{\Phi_{e}}$ yeilds a 512-dimensional multimodal representation $\mathbf{f}_{k}$. Notably, the edge parameters $\Delta h_{ij}$ and $\Delta r_{ij}$ are not embedded at this stage and will be discussed in the following section. 

We train $\mathbf{\Phi_{g}}$ and $\mathbf{\Phi_{e}}$ in a self-supervised manner with three MLP heads: semantic classification $\mathbf{\Theta_{s}}$, bounding box regression $\mathbf{\Theta_{g}}$, and textual similarity $\mathbf{\Theta_{t}}$. The total loss is:
\begin{equation}
       \mathcal{L} = \lambda_{s}\mathcal{L}_{s} + \lambda_{g}\mathcal{L}_{g} + \lambda_{t}\mathcal{L}_{t}.
\end{equation} 
$\mathcal{L}_{s}$ is the cross-entropy loss between $\mathbf{\Theta_{s}}(\mathbf{f}_{k})$ and the semantic label $s_{k}$. $\mathcal{L}_{g}$ combines $\ell_{1}$ and generalized IoU losses for $\mathbf{\Theta_{g}}(\mathbf{f}_{k})$ and $g_{k}$. and $\mathcal{L}_{t}$ is a contrastive loss computed with $\mathbf{\Theta_{t}}(\mathbf{f}_{k})$, $\mathbf{\Phi_{t}}(t^{+}_{k})$, and $\mathbf{\Phi_{t}}(t^{-}_{k})$. $t^{+}_{k}$ and $t^{-}_{k}$ are positive and negative text samples. Hyper-parameters $\lambda_{s}$, $\lambda_{g}$, and $\lambda_{t}$ balance the importance of the modalities, set to 0.5, 1.0, and 0.1 in experiments.

\subsection{Learning Correspondence Matching}
\subsubsection{Objective and Constraints}
We formulate the correspondence matching problem as follows:
\begin{equation}\label{eq:ot}
\begin{aligned}
    &\underset{\mathbf{P} \in \mathbb{R}^{m \times n}}{\mathrm{argmin}} \;\; \sum \mathbf{P} \odot \mathbf{C} + \mathbf{\Omega}(\mathbf{P}, \mathbf{C}), \\
    &\text{subject to} \;\; \mathbf{P} \cdot \mathbf{1}_n = \mathbf{r}, \;\; \mathbf{P}^\mathrm{T} \cdot \mathbf{1}_m = \mathbf{c}.
\end{aligned}
\end{equation}
$m$ and $n$ are the number of matchable nodes in $\mathcal{T}^{A}$ and $\mathcal{T}^{B}$. $\mathbf{P} \in \mathbb{R}^{m \times n}$ is a binary assignment matrix indicating node matches. $\mathbf{C} \in \mathbb{R}^{m \times n}$ is the cost matrix, with lower values for higher similarity. The feasible space of $\mathbf{P}$ is defined by integer vectors $\mathbf{r}$ and $\mathbf{c}$ (lengths $m$ and $n$, values between 0 and 1). $\odot$ denotes the element-wise product, and $\mathbf{\Omega}$ is the regularization function to reduce suboptimal matches. 

To minimize Eq.~\eqref{eq:ot} efficiently, we exploit a novel transformer encoder $\mathbf{\Phi_{m}}$ and an integer programming solver. $\mathbf{\Phi_{m}}$ is equipped with GCA modules to determine unknown matrices $\mathbf{C}$, $\mathbf{r}$, and $\mathbf{c}$. The integer programming solver optimizes the assignment matrix $\mathbf{P}$ constrained by $\mathbf{\Omega}$.

\subsubsection{Grouped Cross-Attention Module}\label{gca}
Unlike previous works \cite{xu2022hierarchical,wu2023screen}, we avoid computing the node-to-node similarity matrix directly. This is due to the quadratic growth of penalty terms in $\mathbf{\Omega}$, which makes Eq.~\eqref{eq:ot} computationally infeasible for large numbers of nodes. Instead, we adopt the divide-and-conquer strategy, performing hierarchical matching from top to bottom. We review the hierarchy of $\mathcal{T}^{B}$ and partition all nodes into several groups based on their association with next-level nodes. Each next-level node and its descendants form a group. This node-to-group matching allows us to recursively estimate correspondences between $\mathcal{T}^{A}$ and $\mathcal{T}^{B}$, significantly reducing computational complexity.

\begin{figure}[t]
  \centering
  \includegraphics[width=1.0\linewidth]{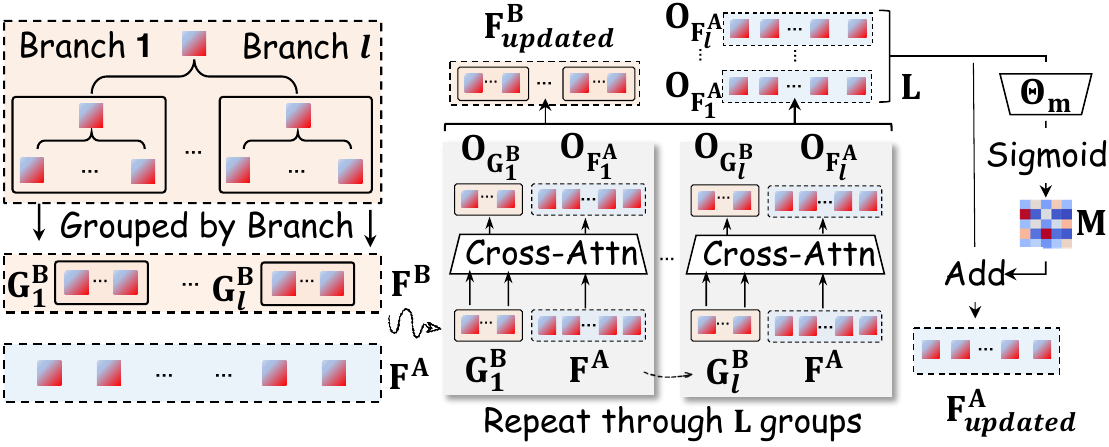}
  \caption{Illustration of the grouped cross-attention module.}
  \label{fig:grouped}
\end{figure}

Taking the first hierarchical matching as an example, trees $\mathcal{T}^{A}$ and $\mathcal{T}^{B}$ are input into frozen first-stage models to obtain 512-dimensional representations: $\mathbf{F}^{A} \in \mathbb{R}^{m \times 512}$ and $\mathbf{F}^{B} \in \mathbb{R}^{n \times 512}$. As shown in Fig.~\ref{fig:grouped}, we assume that $\mathcal{T}^{B}$ has $L$ secondary-level nodes, $\mathbf{F}^{B}$ can be partitioned into $L$ groups, one of which is denoted as $\mathbf{G}^{B}_{l}$ ($1 \le l \le L$). Each group $\mathbf{G}^{B}_{l}$ and the full $\mathbf{F}^{A}$ are iteratively processed by a shared cross-attention module, producing $\mathbf{O}_{\mathbf{G}^{B}_{l}}$ and $\mathbf{O}_{\mathbf{F}^{A}_{l}}$ at the $l^{th}$ iteration. After all iterations, we obtain grouped outputs: $\mathbf{O}_{\mathbf{G}^{B}} \in \mathbb{R}^{n \times 512}$ and $\mathbf{O}_{\mathbf{F}^{A}} \in \mathbb{R}^{L \times m \times 512}$. Using index mapping, $\mathbf{O}_{\mathbf{G}^{B}}$ is added to $\mathbf{F}^{B}$ for updating:
\begin{equation}\label{eq:updateB}
      \mathbf{F}^{B}_{updated} = \mathbf{F}^{B} + \mathbf{O}_{\mathbf{G}^{B}}.
\end{equation} 
In contrast, $\mathbf{O}_{\mathbf{F}^{A}}$ requires additional computation to update $\mathbf{F}^{A}$. We exploit a two-layers MLP head $\mathbf{\Theta_{m}}$ to compute a lower-dimensional matrix $\mathbf{M} \in \mathbb{R}^{L \times m}$ as:
\begin{equation}\label{eq:Matching}
      \mathbf{M} = \mathbf{Sigmoid}(\mathbf{\Theta_{m}}(\mathbf{O}_{\mathbf{F}^{A}})),
\end{equation} 
where $\mathbf{M}$ can implicitly represent the matching probability between $\mathbf{F}^{A}$ and each $\mathbf{G}^{B}_{l}$. This is due to cross-attention, where more relevant $\mathbf{G}^{B}_{l}$ passes more messages into $\mathbf{O}_{\mathbf{F}^{A}_{l}}$, resulting in distinct significance in $\mathbf{O}_{\mathbf{F}^{A}}$ along the first dimension $L$. Consequently, the cost matrix $\mathbf{C}$ in Eq.~\eqref{eq:ot} is obtained as $\mathbf{C} = \mathbf{1} - \mathbf{M}^\top$, and $\mathbf{F}^{A}$ is updated by weighted addition as:
\begin{equation}
\begin{split}
      \mathbf{F}^{A}_{updated} &= \mathbf{F}^{A} + \sum^{L}_{l=1} \mathbf{\hat{M}}_{l}^\top \odot \mathbf{O}_{\mathbf{F}^{A}_{l}},
\end{split}
\end{equation} 
where $\mathbf{\hat{M}} = \frac{\mathbf{M}_{l}}{\sum^{L}_{l=1} \mathbf{M}_{l}}$ is the proportional normalization of $\mathbf{M}$, with higher probabilities yielding higher weights.

Finally, we incorporate GCA modules into the transformer encoder $\mathbf{\Phi_{m}}$ and train it with a binary cross-entropy loss $\mathcal{L}_{m}$. To handle unmatchable cases, we set unmatchable labels as zero-filled vectors during training and filter them out with a threshold $\mathbf{\sigma}=0.5$ during testing. This threshold also helps determine the values of matrix $\mathbf{r}$ and $\mathbf{c}$ in Eq.~\eqref{eq:ot} by comparing the matching probability with $\sigma$.

\subsubsection{Constrained Integer Programming}\label{ip}
Although the matrices $\mathbf{C}$, $\mathbf{r}$, and $\mathbf{c}$ have been specified, the regularization function $\mathbf{\Omega}$ is not determined yet. The purpose of $\mathbf{\Omega}$ is to penalize suboptimal matches that violate either hierarchical or rendering constraints. Specifically, suboptimal matches occur in cases such as: (1) a node $v_i$ that was originally an ancestor of $v_j$ becomes a descendant after matching (i.e., $\Delta h_{ij}$ changes from 0 to 1), or (2) the rendering order between overlapping nodes is reversed (i.e., $\Delta r_{ij}$ changes from 0 to 1). We can define $\mathbf{\Omega}$ as follows:
\begin{equation}\label{eq:regularization}
      \mathbf{\Omega}(\mathbf{P}, \mathbf{C}) = \sum_{\mathbf{P}_{ii'}=1,\mathbf{P}_{jj'}=1} \tau \cdot (\mathbf{C}_{ii'} + \mathbf{C}_{jj'}),
\end{equation} 
where $\mathbf{P}_{ii'}=1$ and $\mathbf{P}_{jj'}=1$ together indicate the suboptimal matches, and $\tau$ controls the penalty strength. Utilizing the edge parameters $\Delta h_{ij}$ and $\Delta r_{ij}$, $\mathbf{\Omega}$ helps maintain consistency in hierarchy and rendering order. 

\subsection{Interactive Construction}
\subsubsection{Universal Data Protocol}
Based on the estimated correspondences, we construct GameUI by integrating the visual attributes of UI nodes into the UX nodes. This involves resizing nodes, updating text content, font, and color, and assigning image assets. To facilitate this integration, we introduce a universal protocol. 

The protocol has two main components: language and compiler. The language defines node entities from low- to high-level semantics, each with attributes such as position, rotation, scale, anchor, opacity, image texture, and text font. The compiler parses this language into various programming languages, enabling cross-platform execution. We use Antlr4 \cite{antlr4}, a powerful parser generator that supports multiple languages, including C++, C\#, Python, and TypeScript. With Antlr4, our protocol can be adapted for game engines like Unity and Unreal Engine, as well as DCC software such as Photoshop and Figma.

\subsubsection{Interactive Web-based Tool}
As shown in Fig.~\ref{fig:teaser}, we implement a web-based interactive tool that takes UI and UX protocols as input and outputs an updated UX protocol with integrated visual attributes. The tool offers several key features: \textbf{Fully-automatic construction} efficiently establishes correspondences and enables pixel-perfect integration from UI design to UX design. \textbf{Semi-automatic revision} assists users in resolving mismatches by highlighting inconsistencies and suggesting potential solutions, primarily through the matrix $\mathbf{M}$ in Eq.~\eqref{eq:Matching}. This enables a human-in-the-loop workflow, allowing users to efficiently identify and address issues based on probabilistic guidance. \textbf{Previewing} offers multiple rendering modes to help users understand and validate designs. \textbf{Editing} allows users to fine-tune the UX design by deleting or creating nodes, repositioning, and updating semantic types. \textbf{Correspondence annotation} helps create a high-quality dataset from real-world games, with cloud backup options. \textbf{State Management} supports session caching and reloading, allowing users to refine GameUI incrementally without losing intermediate progress. These features provide a comprehensive solution for GameUI construction, combining human supervision with machine feedback, ultimately enhancing the overall user experience. 
\begin{figure}[t]
  \centering
  \includegraphics[width=1.0\linewidth]{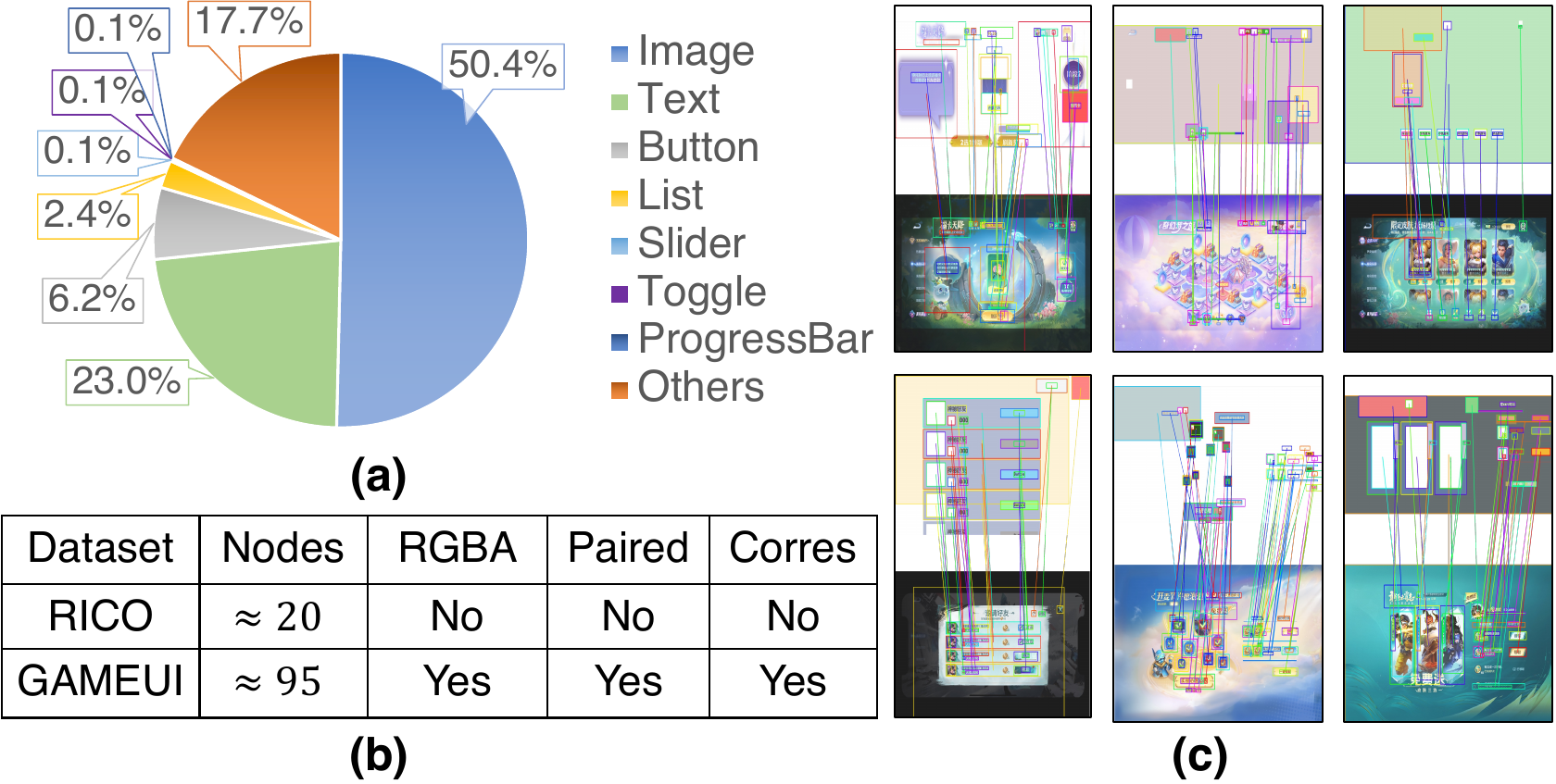}
  \caption{Illustration of the GAMEUI dataset. (a) Distribution of semantic label. (b) Comparison with the RICO dataset. (c) Examples of correspondence annotation.}
  \label{fig:dataset}
\end{figure}

\section{Experiments}
\subsection{Datasets}
To our knowledge, no existing work directly tackles the GameUI construction problem. Most relevant research, such as UI generation or UI retrieval, primarily focuses on the RICO dataset \cite{deka2017rico} or ENRICO dataset \cite{leiva2020enrico}. These public datasets are largely crawled from mobile apps in fields such as communication, medicine, and finance. These datasets are quite simple in aesthetics, lack complex layouts and correspondence annotations. Moreover, their image resources are limited to screenshots instead of renderable RGBA images, making them insufficient for GameUI construction. 

In this paper, we present the GAMEUI dataset (see Fig.~\ref{fig:dataset}), the first dataset designed to build cohesive game UIs from disparate designs. The dataset is collected from different game projects through extensive manual effort by five annotators over 12 months. It comprises 1660 sets of rich, structured game UIs, with each set including a pair of UI and UX designs, along with hierarchy, layout geometry, semantics, and renderable RGBA images. For evaluation, the GAMEUI dataset is split into 1162/166/332 for training, validation, and testing. To avoid overfitting, we adopt comprehensive data augmentation, including random scaling, translation, deletion, and replacement. We also conduct experiments on the RICO dataset, which contains 35851/2109/4218 samples for training, validation, and testing.

\subsection{Implementation Details}
For experiments on the GAMEUI dataset, we adopt a two-stage training strategy (see Fig.~\ref{fig:multimodal_learning}). In the first stage, models are trained for 300 epochs with a batch size of 4 and an initial learning rate of 1e-5, reduced by a factor of 0.2 every 50 epochs. In the second stage, we freeze the first-stage weights and continue training for 200 epochs under the same settings. For the RICO dataset, which lacks correspondence annotations, we only train the first-stage model to learn multimodal representations, with a batch size of 64, an initial learning rate of 1e-4, and 50 epochs to ensure fair comparison with other baselines. We use Google OR-Tools \cite{ortools} to solve the objective in Eq.~\eqref{eq:ot} and set $\tau=1.0$. All experiments are run on an NVIDIA GeForce RTX 3080 GPU with an AMD Ryzen 12-core CPU.
\begin{table*}[t]
    \centering
    \small
 \setlength{\tabcolsep}{4.0mm}{
  \begin{tabular*}{1.0\linewidth}{cccccccccc}
    \toprule
    \textbf{Method} & \textbf{Train Set} & \textbf{HMA} & \textbf{REG} & \textbf{P}$\uparrow$ & \textbf{R}$\uparrow$ & \textbf{F1}$\uparrow$ & \textbf{RMSE}$\downarrow$ & \textbf{PER}$\downarrow$ & \textbf{TIME (s)}$\downarrow$ \\
    \hline
    \multirow{1}{*}{\textbf{LayoutBlend}}      
                                            & \ding{55} & {\ding{55}} & {\ding{55}} & 62.3 & 68.7 & 64.5 & 4.57 & 4.71 & 3.87 \\
    \hline
    \multirow{2}{*}{\textbf{LayoutGMN}}      
                                            & RICO   & {\ding{55}} & {\ding{55}} & 77.1 & 74.4 & 75.3 & 1.65 & 1.99 & \multirow{2}{*}{0.26} \\
                                            & GAMEUI & {\ding{55}} & {\ding{55}} & 81.6 & 81.6 & 81.4 & 1.22 & 0.82 \\
    \hline
    \multirow{2}{*}{\textbf{LayoutTrans}}
                                            & RICO   & {\ding{55}} & {\ding{55}} & 65.4 & 64.4 & 64.4 & 5.55 & 4.91 & \multirow{2}{*}{0.41} \\
                                            & GAMEUI & {\ding{55}} & {\ding{55}} & 71.3 & 68.0 & 68.8 & 2.68 & 1.38 \\
    \hline
    \multirow{2}{*}{\textbf{LayoutDM}}
                                            & RICO   & {\ding{55}} & {\ding{55}} & 67.5 & 65.1 & 65.7 & 3.48 & 3.88 & \multirow{2}{*}{2.95} \\
                                            & GAMEUI & {\ding{55}} & {\ding{55}} & 72.0 & 72.3 & 71.7 & 1.89 & 1.46 \\
    \hline
    \multirow{2}{*}{\textbf{LayoutFlow}}
                                            & RICO   & {\ding{55}} & {\ding{55}} & 81.0 & 80.4 & 80.4 & 2.19 & 1.30 & \multirow{2}{*}{2.48} \\
                                            & GAMEUI & {\ding{55}} & {\ding{55}} & 83.2 & 84.0 & 83.5 & 1.25 & 0.91 \\
    \hline
    \multirow{5}{*}{\textbf{Ours}} 
                                            & RICO   & {\ding{55}} & {\ding{55}} & 82.9 & 83.8 & 83.2 & 1.96 & 1.68 & \multirow{2}{*}{0.76} \\
                                            & GAMEUI & {\ding{55}} & {\ding{55}} & 84.6 & 85.8 & 85.0 & 1.23 & 0.96 \\
                                            & GAMEUI & \ding{51} & {\ding{55}} & 87.7 & 88.3 & 87.6 & 0.81 & 0.44 & 1.01 \\
                                            & GAMEUI & {\ding{55}} & \ding{51} & $\oslash$ & $\oslash$ & $\oslash$ & $\oslash$ & $\oslash$ & $\oslash$ \\
                                            & GAMEUI & \ding{51} & \ding{51} & \textbf{88.2} & \textbf{88.8} & \textbf{88.1} & \textbf{0.21} & \textbf{0.19} & 1.83 \\
     \bottomrule
  \end{tabular*}}
 \caption{Quantitative results on the GAMEUI dataset. \textbf{HMA} and \textbf{REG} denote hierarchical matching and regularization. $\oslash$ indicates a timeout (\textgreater{10s}) caused by applying regularization without hierarchical matching. Note that the second-stage model $\mathbf{\Phi_{m}}$ with GCA modules is exclusively used when hierarchical matching is enabled.}
 \label{tab:evaluation}
\end{table*}

\subsection{Evaluation}
Since there are no directly comparable benchmarks, our comparison was against the most closely related works as: \textbf{LayoutBlend} \cite{xu2022hierarchical} employs a heuristic approach, relying on spatial layout, semantics, and hierarchy to determine node-to-node similarity, and applies the Hungarian algorithm for node correspondence. \textbf{LayoutGMN} \cite{patil2021layoutgmn}, \textbf{LayoutTrans} \cite{gupta2021layouttransformer}, and \textbf{LayoutDM} \cite{inoue2023layoutdm}, and \textbf{LayoutFlow} \cite{guerreiro2024layoutflow} are DNN-based methods that combine spatial layout and semantics for multimodal representation. These DNN models are individually trained on both the RICO and GAMEUI datasets using contrastive, autoregressive, and generative objectives. We use the trained models to extract latent representations of UI and UX nodes, compute node-to-node similarity, and identify the most similar UX node for each UI node as its correspondence.

We evaluate all methods on the GAMEUI dataset, prioritizing three aspects: correspondence matching accuracy, visual consistency of the constructed images, and time efficiency. For matching accuracy, we report precision (\textbf{P}), recall (\textbf{R}), and F1-score (\textbf{F1}) with a weighted macro-average setting. For visual consistency, we integrate the raw RGBA images with correspondences for rendering, and compare the rendered image to the original UI design (ground truth). We adopt Root Mean Squared Error (\textbf{RMSE}) and Pixel-wise Error Ratio (\textbf{PER}) as visual metrics. RMSE measures the average pixel intensity difference (0–255) between the rendered and ground truth images, while PER reports the percentage of pixels with errors. For time efficiency, we record the average time consumption (in seconds) for creating a GameUI from pairwise UI and UX designs. 

\begin{figure}
  \centering
  \includegraphics[width=1.0\linewidth]{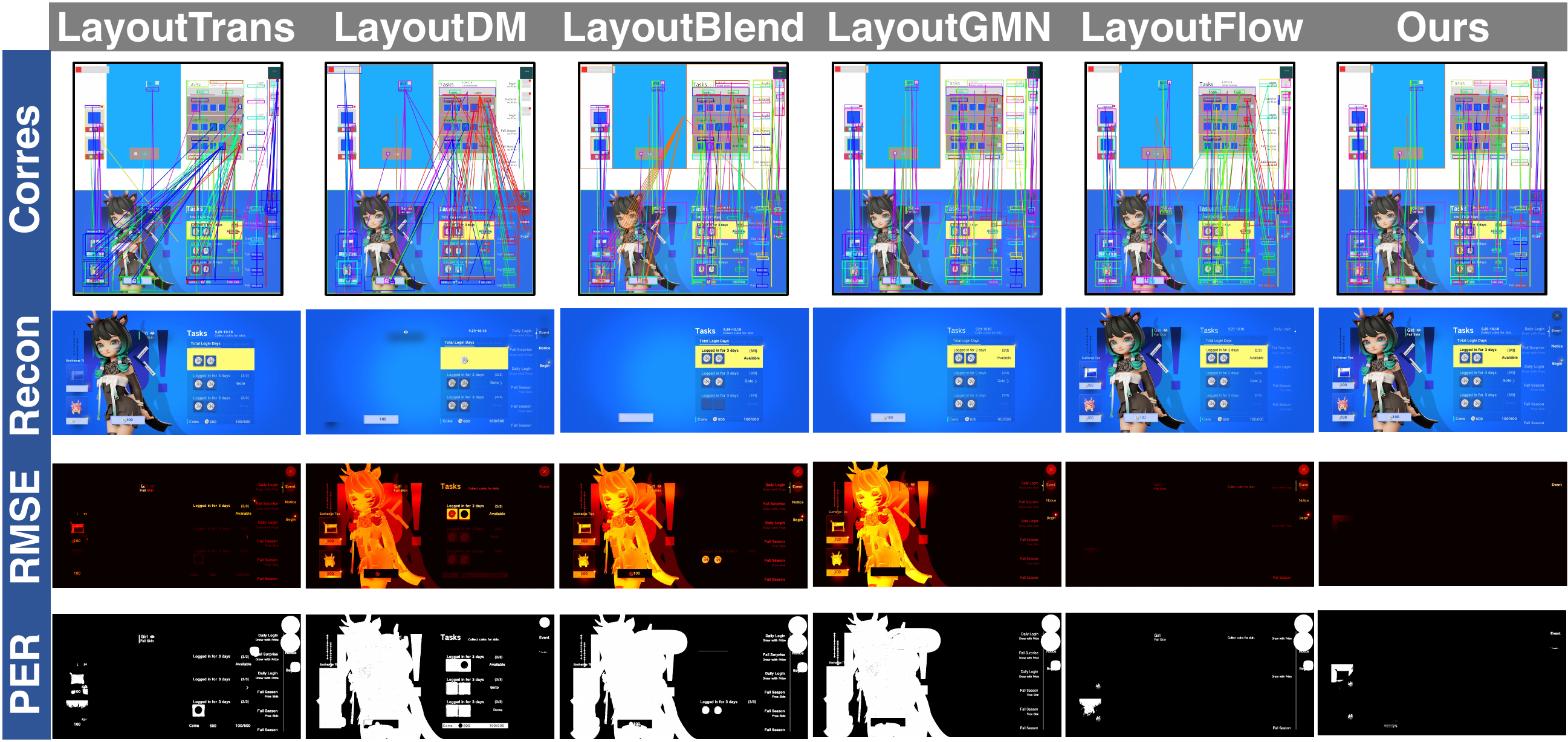}
  \caption{Qualitative results on the GAMEUI dataset.}
  \label{fig:case_study}
\end{figure}
\begin{figure}[t]
  \centering
  \includegraphics[width=1.0\linewidth]{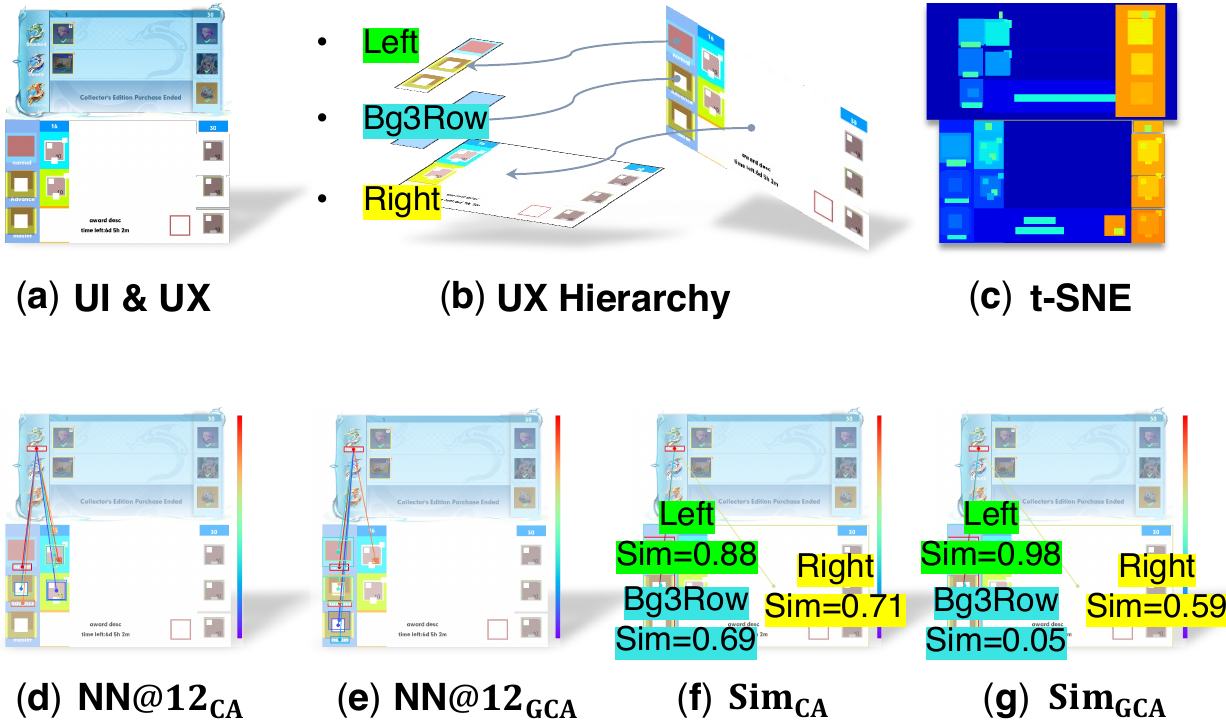}
  \caption{Visualization of the effectiveness of the GCA module. (a, b) show the pairwise UI and UX designs. (c) visualizes the noisy latent representations of UI and UX nodes using t-SNE. (d, e) show the top 12 nearest neighbors of a UI text node in latent space, respectively identified by vanilla cross-attention and GCA. (f, g) show the node-to-group similarity between the UI node and secondary-level UX nodes, with $\mathbf{Sim_{CA}}$ averaged from the node-to-node similarity in vanilla cross-attention.}
  \label{fig:tsne}
\end{figure}
\begin{figure}[t]
  \centering
  \includegraphics[width=1.0\linewidth]{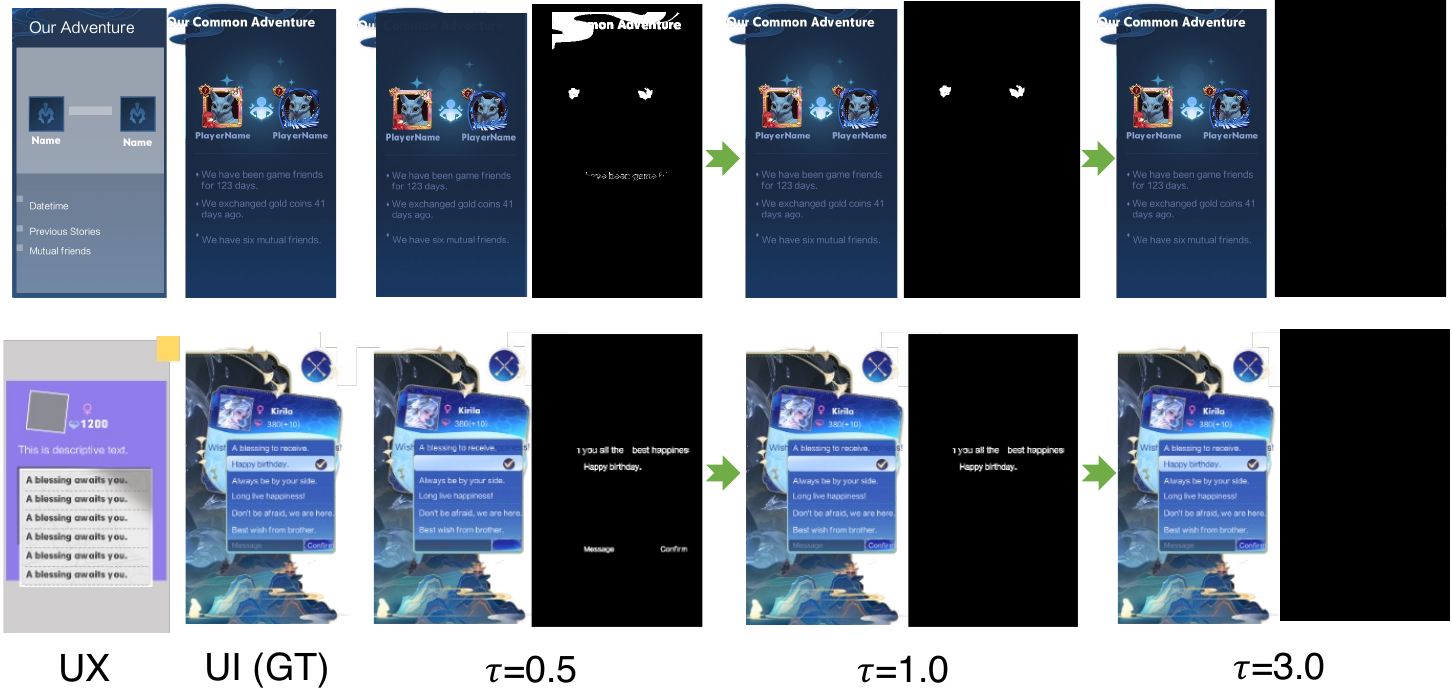}
  \caption{Effectiveness of the regularization function. For each $\tau$, the left shows the construction result and the right shows the PER map measured against the UI design.}
  \label{fig:tau}
\end{figure}

\subsection{Results}\label{Results}
For matching accuracy and visual consistency, Table \ref{tab:evaluation} illustrates that our first and second-stage models outperform other baselines, generalizing well across RICO and GAMEUI. We also compared GAMEUI against the much larger RICO dataset, where the model trained on high-quality GAMEUI performs better on complex game UIs. Fig.~\ref{fig:case_study} qualitatively reflects our superior performance through the correspondences, the constructed image, and the RMSE and PER maps. Regarding time efficiency, our methods remain competitive. The last column of Table \ref{tab:evaluation} shows that hierarchical matching greatly reduces the computational cost introduced by regularization. Without it, most tasks would exceed the time budget and become impractical for real-world use. Although regularization increases runtime, this trade-off is justified by the significant improvements in matching accuracy and visual consistency.

Furthermore, we analyze the shortcomings of baselines, as shown in Table \ref{tab:evaluation} and Fig.~\ref{fig:case_study}. LayoutTrans and LayoutDM perform poorly due to the limited generalization of learnable absolute positional encoding \cite{vaswani2017attention} on complex UIs. In contrast, LayoutFlow’s fixed absolute positional encoding \cite{gehring2017convolutional}, though less flexible, is more stable. LayoutBlend fails on complex UIs due to its naive heuristic rules, while LayoutGMN underperforms because it uses fewer modalities and a simpler training strategy.

Here, we study the GCA module and the regularization function. Fig.~\ref{fig:tsne} reveals that positional and semantic misalignment introduces noise, causing vanilla cross-attention to identify incorrect neighbors. The GCA module addresses this by grouping representations using the UX hierarchy and applying attention within each group, which acts as a filter weighted by hierarchical averaging upon vanilla cross-attention. As a result, the GCA module achieves a better similarity distribution and fewer incorrect neighbors. As illustrated in Fig.~\ref{fig:tau}, we validate the effectiveness of the regularization function, as fewer artifacts in visual consistency as the penalty strength $\tau$ increases. Overall, these strategies progressively improve correspondence matching and visual consistency compared to the base model.

\subsubsection{Ablation Study}
We conduct an ablation study by independently removing input modalities and regularization constraints. Additionally, we replace the GCA module with vanilla cross-attention instead of removing it entirely, to avoid introducing an inductive bias. We can summarize the experimental results in Table~\ref{tab:input-modalities} from three aspects: 

\textbf{Input Modalities.} \textit{Spatial geometry is the most essential modality for our task.} Removing geometry leads to a dramatic drop in F1 score (88.1 to 42.0) and substantial increases in both RMSE (0.21 to 7.03) and PER (0.19 to 5.56).

\textbf{Cross-Attention Modules.} \textit{The GCA module outperforms vanilla cross-attention.} Replacing the GCA module with vanilla cross-attention results in a decrease in F1 (to 86.0) and increases in RMSE (to 1.22) and PER (to 0.82).

\textbf{Regularization Constraints.} \textit{Rendering constraints have a greater impact on performance than hierarchical constraints.} Removing the rendering constraints leads to a larger decrease in F1 (to 86.6) and greater increases in RMSE (to 1.46) and PER (to 0.63), compared to removing the hierarchical constraints (F1 to 87.8, RMSE to 0.43, PER to 0.23). This suggests that rendering constraints play a more important role, especially in maintaining visual consistency.

\begin{table}[t]
  \centering
  \small
  \setlength{\tabcolsep}{3mm}{
  \begin{tabular*}{1.0\linewidth}{lccc}
    \toprule
    \textbf{Ablation} & \textbf{F1}$\uparrow$ & \textbf{RMSE}$\downarrow$ & \textbf{PER}$\downarrow$\\
    \hline
    \multirow{1}{*}{\textbf{Full Model}} & 88.1 & 0.21 & 0.19 \\
    \multirow{1}{*}{$-$ \textit{geometry}} & 42.0 & 7.03 & 5.56 \\
    \multirow{1}{*}{$-$ \textit{semantics}} & 87.2 & 1.05 & 0.59 \\
    \multirow{1}{*}{$-$ \textit{textual content}} & 87.5 & 0.94 & 0.48 \\
    \multirow{1}{*}{$-$ \textit{rendering constraints}} & 86.6 & 1.46 & 0.63 \\
    \multirow{1}{*}{$-$ \textit{hierarchical constraints}} & 87.8 & 0.43 & 0.23 \\
    \multirow{1}{*}{$\dagger$ \textit{vanilla cross-attention}} & 86.0 & 1.22 & 0.82 \\
    \bottomrule
  \end{tabular*}}
  \caption{Ablation study on the proposed model. $-$ represents removing the setting from the full model, and $\dagger$ represents utilizing the setting to replace the GCA module.}
  \label{tab:input-modalities}
\end{table}

\begin{figure}[t]
  \centering
  \includegraphics[width=1.0\linewidth]{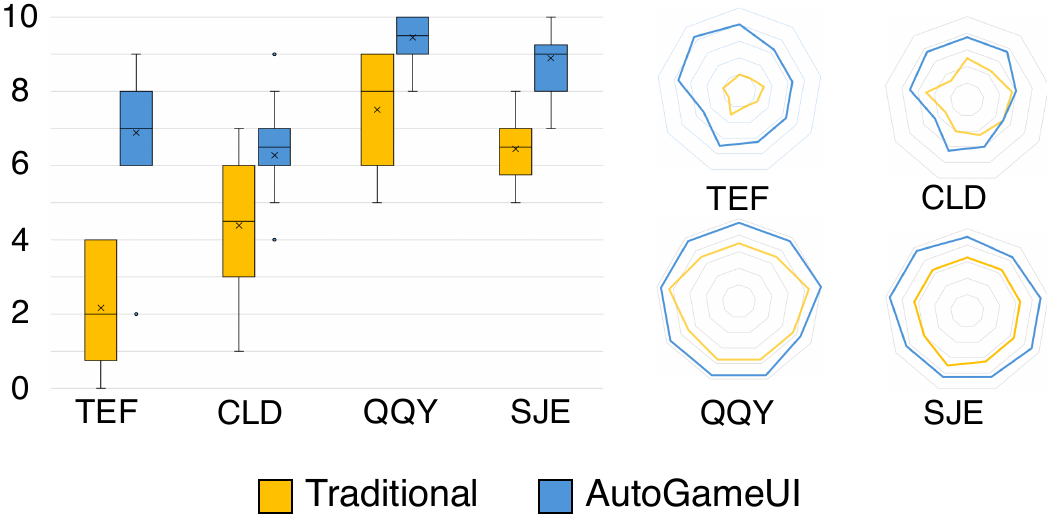}
  \caption{The statistical data of the expert evaluation by comparing our system with the traditional manual workflow.}
  \label{fig:user_study}
\end{figure}

\subsubsection{Expert Evaluation}\label{User}
To evaluate our system with realistic user feedback, we designed an expert-driven study comparing it with the traditional manual workflow. Following common human-centric practices in UI automation, we defined four metrics: Time Efficiency (\textbf{TEF}, normalized time statistics, e.g. $\mathbf{TEF}_{trad}=Time_{ours}/(Time_{ours}+Time_{trad}) \times 10$), Cognitive Load (\textbf{CLD}, perceived mental effort), Output Quality (\textbf{OQY}, perceptible visual fidelity), and Subjective Experience (\textbf{SJE}, overall satisfaction and ease of use). Except for TEF, all metrics were scored on a 0-10 scale, with five levels: Unsatisfactory (0-2), Needs Improvement (2-4), Satisfactory (4-6), Good (6-8), and Excellent (8-10).

Nine experts (5 men, 4 women) were invited to participate in the study. All were familiar with the traditional workflow and had at least five years of professional experience. Each expert was assigned two distinct UI/UX pairs, randomly shuffled to mitigate sequence bias. They used identical desktop PCs and received the same training in scoring criteria and system usage. They could choose which method to try first, but had to complete each task before switching.

The results of this study are shown in Fig.~\ref{fig:user_study}. The box plot on the left shows that our system outperforms the traditional workflow across all four metrics, with higher medians and less variability. Specifically, time efficiency improves by up to 3$\times$, and outputs are more visually accurate with few artifacts. The radar charts on the right provide a view to per-user preference, showing that our system achieves consistently high scores among all users, while the traditional workflow exhibits generally lower ratings.

\section{Conclusion}
We present a novel system for constructing cohesive game
user interfaces, supporting both automatic and interactive
pipelines. The automatic pipeline employs a two-stage multimodal learning method to match UI and UX designs, while the interactive pipeline provides multiple assistant features for a user-friendly experience. Extensive experiments on our newly built GAMEUI dataset and the RICO dataset demonstrate the effectiveness. Our system has been deployed in several mobile games, delivering significant improvements.

\section{Acknowledgments}
We gratefully acknowledge the continuous support provided by our colleagues Jun Deng and Fei Ling. Their contributions in terms of project guidance and funding have been invaluable to the completion of this work. 

\bibliography{aaai2026}

@inproceedings{he2021actionbert,
  title={Actionbert: Leveraging user actions for semantic understanding of user interfaces},
  author={He, Zecheng and Sunkara, Srinivas and Zang, Xiaoxue and Xu, Ying and Liu, Lijuan and Wichers, Nevan and Schubiner, Gabriel and Lee, Ruby and Chen, Jindong},
  booktitle={Proceedings of the AAAI Conference on Artificial Intelligence},
  volume={35},
  number={7},
  pages={5931--5938},
  year={2021}
}

@inproceedings{kumar2011flexible,
  title={Flexible tree matching},
  author={Kumar, Ranjitha and Talton, Jerry O and Ahmad, Salman and Roughgarden, Tim and Klemmer, Scott R},
  booktitle={Proceedings of the Twenty-Second International Joint Conference on Artificial Intelligence-Volume Volume Three},
  pages={2674--2679},
  year={2011}
}

@inproceedings{kumar2011bricolage,
  title={Bricolage: example-based retargeting for web design},
  author={Kumar, Ranjitha and Talton, Jerry O and Ahmad, Salman and Klemmer, Scott R},
  booktitle={Proceedings of the SIGCHI Conference on Human Factors in Computing Systems},
  pages={2197--2206},
  year={2011}
}

@inproceedings{bacikova2013defining,
  title={Defining domain language of graphical user interfaces},
  author={Bacikov{\'a}, Michaela and Porub{\"a}n, Jaroslav and Lakatos, Dominik},
  booktitle={2nd Symposium on Languages, Applications and Technologies (2013)},
  year={2013},
  organization={Schloss-Dagstuhl-Leibniz Zentrum f{\"u}r Informatik}
}

@article{vaswani2017attention,
  title={Attention is all you need},
  author={Vaswani, Ashish},
  journal={arXiv preprint arXiv:1706.03762},
  year={2017}
}

@inproceedings{gehring2017convolutional,
  title={Convolutional sequence to sequence learning},
  author={Gehring, Jonas and Auli, Michael and Grangier, David and Yarats, Denis and Dauphin, Yann N},
  booktitle={International conference on machine learning},
  pages={1243--1252},
  year={2017},
  organization={PMLR}
}

@inproceedings{deka2017rico,
  title={Rico: A mobile app dataset for building data-driven design applications},
  author={Deka, Biplab and Huang, Zifeng and Franzen, Chad and Hibschman, Joshua and Afergan, Daniel and Li, Yang and Nichols, Jeffrey and Kumar, Ranjitha},
  booktitle={Proceedings of the 30th annual ACM symposium on user interface software and technology},
  pages={845--854},
  year={2017}
}

@article{vinyals2019grandmaster,
  title={Grandmaster level in StarCraft II using multi-agent reinforcement learning},
  author={Vinyals, Oriol and Babuschkin, Igor and Czarnecki, Wojciech M and Mathieu, Micha{\"e}l and Dudzik, Andrew and Chung, Junyoung and Choi, David H and Powell, Richard and Ewalds, Timo and Georgiev, Petko and others},
  journal={nature},
  volume={575},
  number={7782},
  pages={350--354},
  year={2019},
  publisher={Nature Publishing Group}
}

@inproceedings{ye2020mastering,
  title={Mastering complex control in moba games with deep reinforcement learning},
  author={Ye, Deheng and Liu, Zhao and Sun, Mingfei and Shi, Bei and Zhao, Peilin and Wu, Hao and Yu, Hongsheng and Yang, Shaojie and Wu, Xipeng and Guo, Qingwei and others},
  booktitle={Proceedings of the AAAI Conference on Artificial Intelligence},
  volume={34},
  number={04},
  pages={6672--6679},
  year={2020}
}

@inproceedings{dayama2021interactive,
  title={Interactive layout transfer},
  author={Dayama, Niraj Ramesh and Santala, Simo and Br{\"u}ckner, Lukas and Todi, Kashyap and Du, Jingzhou and Oulasvirta, Antti},
  booktitle={Proceedings of the 26th International Conference on Intelligent User Interfaces},
  pages={70--80},
  year={2021}
}

@inproceedings{yamaguchi2021canvasvae,
  title={Canvasvae: Learning to generate vector graphic documents},
  author={Yamaguchi, Kota},
  booktitle={Proceedings of the IEEE/CVF International Conference on Computer Vision},
  pages={5481--5489},
  year={2021}
}

@inproceedings{li2021screen2vec,
  title={Screen2vec: Semantic embedding of gui screens and gui components},
  author={Li, Toby Jia-Jun and Popowski, Lindsay and Mitchell, Tom and Myers, Brad A},
  booktitle={Proceedings of the 2021 CHI Conference on Human Factors in Computing Systems},
  pages={1--15},
  year={2021}
}

@inproceedings{gupta2021layouttransformer,
  title={Layouttransformer: Layout generation and completion with self-attention},
  author={Gupta, Kamal and Lazarow, Justin and Achille, Alessandro and Davis, Larry S and Mahadevan, Vijay and Shrivastava, Abhinav},
  booktitle={Proceedings of the IEEE/CVF International Conference on Computer Vision},
  pages={1004--1014},
  year={2021}
}

@inproceedings{inoue2023layoutdm,
  title={Layoutdm: Discrete diffusion model for controllable layout generation},
  author={Inoue, Naoto and Kikuchi, Kotaro and Simo-Serra, Edgar and Otani, Mayu and Yamaguchi, Kota},
  booktitle={Proceedings of the IEEE/CVF Conference on Computer Vision and Pattern Recognition},
  pages={10167--10176},
  year={2023}
}

@inproceedings{guerreiro2024layoutflow,
  title={Layoutflow: flow matching for layout generation},
  author={Guerreiro, Julian Jorge Andrade and Inoue, Naoto and Masui, Kento and Otani, Mayu and Nakayama, Hideki},
  booktitle={European Conference on Computer Vision},
  pages={56--72},
  year={2024},
  organization={Springer}
}

@article{fan2022minedojo,
  title={Minedojo: Building open-ended embodied agents with internet-scale knowledge},
  author={Fan, Linxi and Wang, Guanzhi and Jiang, Yunfan and Mandlekar, Ajay and Yang, Yuncong and Zhu, Haoyi and Tang, Andrew and Huang, De-An and Zhu, Yuke and Anandkumar, Anima},
  journal={Advances in Neural Information Processing Systems},
  volume={35},
  pages={18343--18362},
  year={2022}
}

@article{batifol2025flux,
  title={FLUX. 1 Kontext: Flow Matching for In-Context Image Generation and Editing in Latent Space},
  author={Batifol, Stephen and Blattmann, Andreas and Boesel, Frederic and Consul, Saksham and Diagne, Cyril and Dockhorn, Tim and English, Jack and English, Zion and Esser, Patrick and Kulal, Sumith and others},
  journal={arXiv e-prints},
  pages={arXiv--2506},
  year={2025}
}

@article{zhao2025hunyuan3d,
  title={Hunyuan3d 2.0: Scaling diffusion models for high resolution textured 3d assets generation},
  author={Zhao, Zibo and Lai, Zeqiang and Lin, Qingxiang and Zhao, Yunfei and Liu, Haolin and Yang, Shuhui and Feng, Yifei and Yang, Mingxin and Zhang, Sheng and Yang, Xianghui and others},
  journal={arXiv preprint arXiv:2501.12202},
  year={2025}
}

@inproceedings{takada2023genelive,
  title={Gen{\'e}live! generating rhythm actions in love live!},
  author={Takada, Atsushi and Yamazaki, Daichi and Yoshida, Yudai and Ganbat, Nyamkhuu and Shimotomai, Takayuki and Hamada, Naoki and Liu, Likun and Yamamoto, Taiga and Sakurai, Daisuke},
  booktitle={Proceedings of the AAAI Conference on Artificial Intelligence},
  volume={37},
  number={4},
  pages={5266--5275},
  year={2023}
}

@book{antlr4,
  title={The definitive ANTLR 4 reference},
  author={Parr, Terence},
  year={2013},
  publisher={Pragmatic Bookshelf},
  url={https://www.antlr.org}
}

@article{anastassiou2024seed,
  title={Seed-TTS: A Family of High-Quality Versatile Speech Generation Models},
  author={Anastassiou, Philip and Chen, Jiawei and Chen, Jitong and Chen, Yuanzhe and Chen, Zhuo and Chen, Ziyi and Cong, Jian and Deng, Lelai and Ding, Chuang and Gao, Lu and others},
  journal={arXiv preprint arXiv:2406.02430},
  year={2024}
}

@article{liu2023emage,
  title={Emage: Towards unified holistic co-speech gesture generation via masked audio gesture modeling},
  author={Liu, Haiyang and Zhu, Zihao and Becherini, Giorgio and Peng, Yichen and Su, Mingyang and Zhou, You and Iwamoto, Naoya and Zheng, Bo and Black, Michael J},
  journal={arXiv preprint arXiv:2401.00374},
  year={2023}
}

@inproceedings{chen2024taming,
  title={Taming Diffusion Probabilistic Models for Character Control},
  author={Chen, Rui and Shi, Mingyi and Huang, Shaoli and Tan, Ping and Komura, Taku and Chen, Xuelin},
  booktitle={ACM SIGGRAPH 2024 Conference Papers},
  pages={1--10},
  year={2024}
}

@inproceedings{reimers-2019-sentence-bert,
  title = "Sentence-BERT: Sentence Embeddings using Siamese BERT-Networks",
  author = "Reimers, Nils and Gurevych, Iryna",
  booktitle = "Proceedings of the 2019 Conference on Empirical Methods in Natural Language Processing",
  month = "11",
  year = "2019",
  publisher = "Association for Computational Linguistics",
  url = "https://arxiv.org/abs/1908.10084",
}

@inproceedings{leiva2020enrico,
  title={Enrico: A dataset for topic modeling of mobile UI designs},
  author={Leiva, Luis A and Hota, Asutosh and Oulasvirta, Antti},
  booktitle={22nd International Conference on Human-Computer Interaction with Mobile Devices and Services},
  pages={1--4},
  year={2020}
}

@inproceedings{patil2021layoutgmn,
  title={Layoutgmn: Neural graph matching for structural layout similarity},
  author={Patil, Akshay Gadi and Li, Manyi and Fisher, Matthew and Savva, Manolis and Zhang, Hao},
  booktitle={Proceedings of the IEEE/CVF Conference on Computer Vision and Pattern Recognition},
  pages={11048--11057},
  year={2021}
}

@inproceedings{wu2021screen,
  title={Screen parsing: Towards reverse engineering of ui models from screenshots},
  author={Wu, Jason and Zhang, Xiaoyi and Nichols, Jeff and Bigham, Jeffrey P},
  booktitle={The 34th Annual ACM Symposium on User Interface Software and Technology},
  pages={470--483},
  year={2021}
}

@inproceedings{li2021understanding,
  title={Understanding players’ interaction patterns with mobile game app UI via visualizations},
  author={Li, Quan and Zeng, Haipeng and Peng, Zhenhui and Ma, Xiaojuan},
  booktitle={Proceedings of the Ninth International Symposium of Chinese CHI},
  pages={9--21},
  year={2021}
}

@inproceedings{gonccalves2023my,
  title={" My Zelda Cane": Strategies Used by Blind Players to Play Visual-Centric Digital Games},
  author={Gon{\c{c}}alves, David and Pi{\c{c}}arra, Manuel and Pais, Pedro and Guerreiro, Jo{\~a}o and Rodrigues, Andr{\'e}},
  booktitle={Proceedings of the 2023 CHI conference on human factors in computing systems},
  pages={1--15},
  year={2023}
}

@article{xu2022hierarchical,
  title={Hierarchical layout blending with recursive optimal correspondence},
  author={Xu, Pengfei and Li, Yifan and Yang, Zhijin and Shi, Weiran and Fu, Hongbo and Huang, Hui},
  journal={ACM Transactions on Graphics (TOG)},
  volume={41},
  number={6},
  pages={1--15},
  year={2022},
  publisher={ACM New York, NY, USA}
}

@inproceedings{li2022learning,
  title={Learning to denoise raw mobile UI layouts for improving datasets at scale},
  author={Li, Gang and Baechler, Gilles and Tragut, Manuel and Li, Yang},
  booktitle={Proceedings of the 2022 CHI Conference on Human Factors in Computing Systems},
  pages={1--13},
  year={2022}
}

@inproceedings{warner2023interactive,
  title={Interactive Flexible Style Transfer for Vector Graphics},
  author={Warner, Jeremy and Kim, Kyu Won and Hartmann, Bjoern},
  booktitle={Proceedings of the 36th Annual ACM Symposium on User Interface Software and Technology},
  pages={1--14},
  year={2023}
}

@article{wu2023screen,
  title={Screen correspondence: Mapping interchangeable elements between uis},
  author={Wu, Jason and Swearngin, Amanda and Zhang, Xiaoyi and Nichols, Jeffrey and Bigham, Jeffrey P},
  journal={arXiv preprint arXiv:2301.08372},
  year={2023}
}

@inproceedings{bai2023layout,
  title={Layout representation learning with spatial and structural hierarchies},
  author={Bai, Yue and Manandhar, Dipu and Wang, Zhaowen and Collomosse, John and Fu, Yun},
  booktitle={Proceedings of the AAAI Conference on Artificial Intelligence},
  volume={37},
  number={1},
  pages={206--214},
  year={2023}
}

@article{panayiotou2024smauto,
  title={SmAuto: A domain-specific-language for application development in smart environments},
  author={Panayiotou, Konstantinos and Doumanidis, Constantine and Tsardoulias, Emmanouil and Symeonidis, Andreas L},
  journal={Pervasive and Mobile Computing},
  volume={101},
  pages={101931},
  year={2024},
  publisher={Elsevier}
}

@article{su2024roformer,
  title={Roformer: Enhanced transformer with rotary position embedding},
  author={Su, Jianlin and Ahmed, Murtadha and Lu, Yu and Pan, Shengfeng and Bo, Wen and Liu, Yunfeng},
  journal={Neurocomputing},
  volume={568},
  pages={127063},
  year={2024},
  publisher={Elsevier}
}

@inproceedings{jiang2024graph4gui,
  title={Graph4GUI: Graph Neural Networks for Representing Graphical User Interfaces},
  author={Jiang, Yue and Zhou, Changkong and Garg, Vikas and Oulasvirta, Antti},
  booktitle={Proceedings of the CHI Conference on Human Factors in Computing Systems},
  pages={1--18},
  year={2024}
}

@article{otani2024ltsim,
  title={LTSim: Layout Transportation-based Similarity Measure for Evaluating Layout Generation},
  author={Otani, Mayu and Inoue, Naoto and Kikuchi, Kotaro and Togashi, Riku},
  journal={arXiv preprint arXiv:2407.12356},
  year={2024}
}

@misc{ortools,
  author       = {Laurent Perron and Vincent Furnon},
  year         = {2024},
  title        = {Route. Schedule. Plan. Assign. Pack. Solve.},
  howpublished = {\url{https://developers.google.com/optimization/}},
  note         = {Accessed: 2025-11-27}
}

@misc{uizard,
  author       = {Uizard},
  year         = {2024},
  title        = {Uizard: UI Design Made Easy, Powered By AI},
  howpublished = {\url{https://uizard.io}},
  note         = {Accessed: 2025-11-27}
}

@misc{framer,
  author       = {Framer},
  year         = {2025},
  title        = {Framer: Create a professional website, free. No code website builder loved by designers.},
  howpublished = {\url{https://www.framer.com}},
  note         = {Accessed: 2025-11-27}
}

\end{document}